\documentstyle[aps,prb,multicol,epsf]{revtex}

\begin{document}
\draft
\preprint{}
\wideabs{
\title{Electronic States and Excitation Spectra of Copper Oxides with Ladder
 and/or Chain}

\author{Yoshiaki Mizuno$^{\rm a,b}$, Takami Tohyama$^{\rm b}$
        and 
        Sadamichi Maekawa$^{\rm b}$
       }
\address{$^{\rm a}$Department of Applied Physics, 
                                Nagoya University, Nagoya 464-01, Japan}
\address{$^{\rm b}$Institute for Materials Research, 
                               Tohoku University, Sendai 980-77, Japan}

\maketitle
\begin{abstract}
Recently the superconductivity has been reported in the copper oxides (Sr,Ca)$_{14}$Cu$_{24}$O$_{41}$ with ladders and edge-sharing chains under pressure.  In order to understand the mechanism of the superconductivity, it is crucial to clarify the distribution of self-doped holes. From the analysis of the optical conductivity by ionic model and exact diagonalization method, we confirm that with substituting Ca for Sr, the self-doped holes are transferred from the chains to the ladders. We also examine the electronic states of edge-sharing chain in a variety of copper oxides, and derive the effective Hamiltonian. The dependences of the magnetic interactions between Cu ions and of the hopping energies of the Zhang-Rice singlet on the bond angle of Cu-O-Cu are discussed.
\end{abstract}
}
\narrowtext

\section{Introduction}
Spin ladder system with two legs has drawn much attention, because of the possibility of the superconductivity as predicted theoretically. Recently, the superconductivity in doped spin ladders (Sr,Ca)$_{14}$Cu$_{24}$O$_{41}$ under high pressure has been reported\cite{Uehara}, and its relation to high-{\it T}$_{\rm c}$ superconductivity has been discussed. In (Sr,Ca)$_{14}$Cu$_{24}$O$_{41}$, there are not only ladders but also edge-sharing chains. In order to understand the electronic structure of (Sr,Ca)$_{14}$Cu$_{24}$O$_{41}$, it is important to examine the nature of edge-sharing Cu-O chain. In the previous study, we have discussed the electronic states of ladder for (Sr,Ca)$_{14}$Cu$_{24}$O$_{41}$\cite{Mizuno}.  In this paper, we examine theoretically the electronic states, the excitation spectra and magnetic interactions of the edge-sharing chains for (Sr,Ca)$_{14}$Cu$_{24}$O$_{41}$.  In this type of chain, the strength and sign of the magnetic interactions are expected to be changed by the bond angle $\theta$ of Cu-O-Cu.  We also present a systematic investigation of the exchange interactions and the hopping energies for the Zhang-Rice singlet for various materials with edge-sharing structure.

\section{The electronic states of Sr$_{\rm 14-x}$Ca$_{\rm x}$Cu$_{24}$O$_{41}$}
This material consists of the alternating stacks of the layer of edge-sharing chains, the layer of  two-leg ladders, and (Sr,Ca) layer. The average valence of copper is +2.25, so that the material is a self-doped system. With the substitution of Ca for Sr, the resistivity changes from the semi-conductive behavior to the metallic one, and the superconductivity occurs under pressure. Therefore, it is significant to know the change of distribution of the self-doped holes with Ca doping. In our previous study, it has been found from the analysis of Madelung energy that with increasing Ca content holes on the ladder is more stable that on the chain\cite{Mizuno}. 

In this section, we focus on the electronic states of edge-sharing chain, and discuss the distribution of holes from the viewpoint on effect of hole doping on the electronic states of the edge-sharing chain.

We perform theoretical investigation on the optical conductivity $\sigma(\omega)$ for the edge-sharing chain by the ionic and cluster model approaches. The present materials are classified as a charge transfer (CT) type. The CT energy $\Delta$ is defined according to the prescription of the previous study\cite{Mizuno,Ohta}.  The hopping parameters are determined using the relation by Slater and Koster\cite{Slater}.  We assume the bond length dependences with $d^{-4}$ and $d^{-3}$ for hopping integrals $(pd\sigma)$ and $(pp\sigma)$, respectively.\cite{parameter} The on-site Coulomb energies are set to be $U_d$=8.5 eV and $U_p$=4.1~eV as for the previous studies.\cite{Ohta}  The Hund's coupling on O ion is assumed to be $K_p$=0.6~eV, and the direct exchange interaction between Cu3$d$ and O2$p$ is taken to be $K_{pd}$=0.05~eV, which gives the ferromagnetic interaction between nearest neighbor Cu spins. The exact diagonalization method is used for the calculation of $\sigma(\omega)$ for the Cu$_4$O$_{10}$ cluster with free boundary condition. 

The results for $\sigma(\omega)$ are shown in Fig. 1. In the non-doped case (a), two main structures {\bf A} and {\bf B} are observed. The structure {\bf A} is due to the CT excitation, determining the CT gap. The intensity is weak because the hole propagation is suppressed due to the near 90$^\circ$ edge-sharing structure. On the other hand, the structure {\bf B} is related to a local excitation with CuO$_4$, and has strong intensity. 

In the holes doped case (b), delocalized holes produce the low-energy Drude weight at $\omega$=0.5 eV, which may contribute to the metallic behavior. The important point is that a new structure {\bf C} appears at around 3 eV between {\bf A} and {\bf B} structures, and with introducing more holes  (c) its intensity becomes stronger. This structure is caused by a local excitation from non-bonding state of oxygen to the Zhang-Rice (ZR) local singlet state. Therefore, if holes are transferred from the chain to the ladder with Ca substitution, this intensity decreases. This excitation exists even in the ladder and CuO$_2$ plane, but it is difficult to observe it due to overlapping with the large structure related to CT excitation. 

In the experiment, in Sr$_{14}$Cu$_{24}$O$_{41}$ a structure with strong intensity is observed at around 3 eV, and decreases with Ca doping\cite{Osafune}. This intensity is attributed to the excitation of chain, since $\sigma(\omega)$ for ladder does not give any predominant structures at the region\cite{Mizuno}.  The decrease in the intensity with increasing Ca content is consistent with theoretical results and is an evidence that Ca substitution leads to transfer of holes from chain to ladder.

\begin{figure}
\begin{center}
\epsfile{file=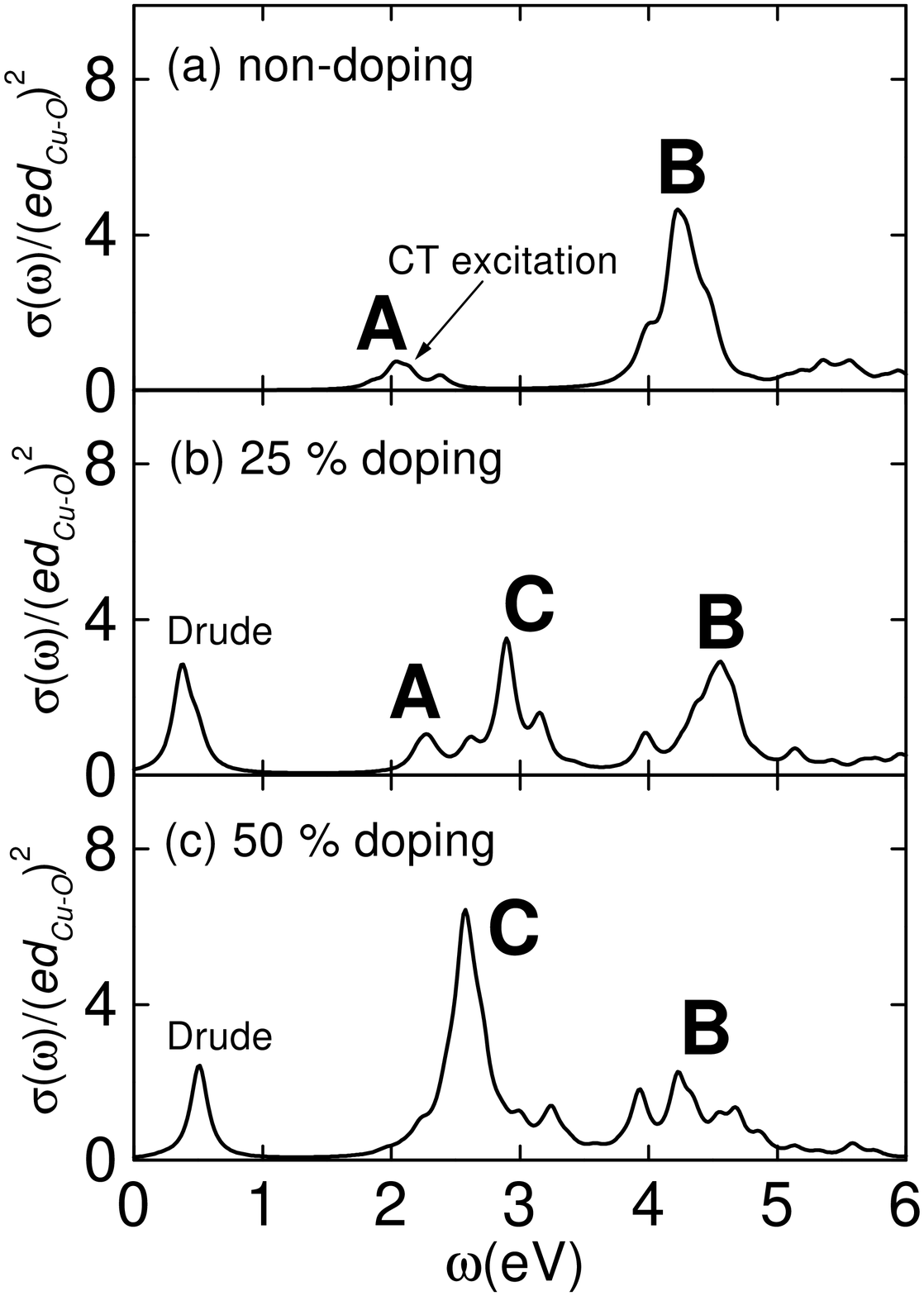,scale=0.36}
\caption{Optical conductivity $\sigma(\omega)$ for the Cu$_{4}$O$_{10}$ cluster. (a) non-doping case, (b) 25 $\%$ doping case and (c) 50 $\%$ doping case. $d_{\rm Cu-O}$ and $e$ are the bond length between Cu and O and the elementary electric charge,  respectively, and $\hbar = c = 1$. The $\delta$ functions are convoluted with a Lorentzian broadening of 0.08~eV.}
\label{fig1}
\end{center}
\end{figure}

\section{The magnetic interactions and hopping energies} 
In this section, the magnetic interactions between Cu ions and the hopping energies of the ZR local singlet are examined for several materials with edge-sharing chain. The materials are La$_6$Ca$_8$Cu$_{24}$O$_{41}$, Li$_2$CuO$_2$, CuGeO$_3$ and Ca$_2$Y$_2$Cu$_5$O$_{10}$. In the edge-sharing chain, there exist nearest neighbor  ferromagentic (FM) or antiferromagnetic (AFM) exchange interactions ($J_1$) and second nearest neighbor AFM exchange interactions ($J_2$). 

We calculate $J_1$ and $J_2$ by using the exact diagonalization method for small clusters\cite{Mizuno2}.  The parameter values are determined by the same procedure used in the calculation of $\sigma(\omega)$.  $J_1$ and $J_2$ can be defined by the energy difference between the lowest singlet and the lowest triplet states in Cu$_2$O$_6$ and Cu$_2$O$_8$ clusters with two holes, respectively. 

The results for the magnetic interactions are shown in Fig.~2(a). The positive $J_1$ is AFM, while the negative one is FM. In order to examine their dependences on $\theta$, we calculate $J_1$ and $J_2$ for several choices of parameter sets of $d_{\rm Cu-O}$ and $\Delta$.   The thick and thin solid lines denote the results for $d_{\rm Cu-O}$=1.90\AA\ and 1.97\AA, respectively, taking $\Delta$=3.0~eV as a typical value for the 1D cuprates.  $d_{\rm Cu-O}$=1.90\AA\ and 1.97\AA\  correspond to the lower and upper bounds of $d_{\rm Cu-O}$ for the cuprates, respectively.  The increase of $d_{\rm Cu-O}$ causes the decrease of hopping amplitudes, resulting in the reduction of the magnitudes of $J_1$ and $J_2$.  The dotted line denotes the result for $d_{\rm Cu-O}$=1.95\AA\ and $\Delta$=5.0~eV.  The large $\Delta$ also reduces the magnitudes of $J_1$ and $J_2$ as expected.  The bond angle minimizing $J_1$ deviates from 90 degrees.  This is due to AFM contribution from direct O-O hoppings. In contrast to $J_1$, $J_2$ decreases slightly with increasing $\theta$.   This stems from the increase of $d_{\rm O-O}$ along the chains. 

By using $J_1$ and $J_2$ obtained from small cluster, the experiment of the susceptibility is in accord with the theory for Li$_2$CuO$_2$, La$_6$Ca$_8$Cu$_{24}$O$_{41}$ and Ca$_2$Y$_2$Cu$_5$O$_{10}$, but is not for CuGeO$_3$. In fact, O ions in chain are strongly coupled to Ge ions\cite{Khomskii}, and (2) the change of the lattice constant perpendicular to the chain due to temperature is larger than that of the others\cite{Fujita}. We think that if these effects are taken into account, it is possible to explain the temperature dependence of susceptibility.

The 90$^\circ$ Cu-O-Cu bond structures are also seen in ladder materials. Those bond structures give interladder interactions. We can calculate these interactions, $J_{\rm inter}$ as well as $J_1$. The calculated values of $J_{\rm inter}$ for several materials are shown with the solid symbols in Fig. 2(a). $J_{\rm inter}$'s are FM and $\sim$-300 K. These interactions are not small as compared with intraladder interactions ($\sim$1500 K). Therefore, we should include them in the study of the magnetic properties.

Next, we estimate the hopping parameters $t_1$ and $t_2$, which are the hopping energies of the ZR local singlet between nearest neighbor and second nearest neighbor CuO$_4$ units, respectively\cite{ef}. $t_1$ and $t_2$ are obtained from the energy differences between the ground and first excited  states in Cu$_2$O$_6$ and Cu$_2$O$_8$ clusters with 3 holes, respectively.

The several lines in Fig. 2(b) show the $\theta$ dependence of $t_1$ and $t_2$ for the same parameter sets as Fig. 2(a). With increasing $\theta$ from 80$^\circ$ to 100$^\circ$, $t_1$ decreases from 500 K to -2000 K, changing the sign from positive to negative. On the contrary, the changes of $t_2$ are small, and its sign remains unchanged. 
Reflecting the 90$^\circ$ bonding, $t_1$ is smaller than that for the 180$^\circ$ bonding ($\sim$ 3500~K).
It is interesting to note that in $\theta>$94$^\circ$ or $\theta<$94$^\circ$ the strength of $|t_1|$ is larger or smaller than that of $|t_2|$, respectively. This suggests that the hole motion in the edge-sharing chain is sensitive to $\theta$.  As one can see in Fig. 2(b), $|t_1|<|t_2|$ for La$_6$Ca$_8$Cu$_{24}$O$_{41}$, $|t_1|\sim|t_2|$ for Li$_2$CuO$_2$, and $|t_1|>|t_2|$ for CuGeO$_{3}$. The interladder hopping energies $t_{\rm inter}$ are $\sim$-600K and $|t_{\rm inter}/t_{\rm intra}|\sim0.16$. These hopping energies are important to examine the angle-resolved photoemission (ARPES) spectra. 

\begin{figure}
\begin{center}
\epsfile{file=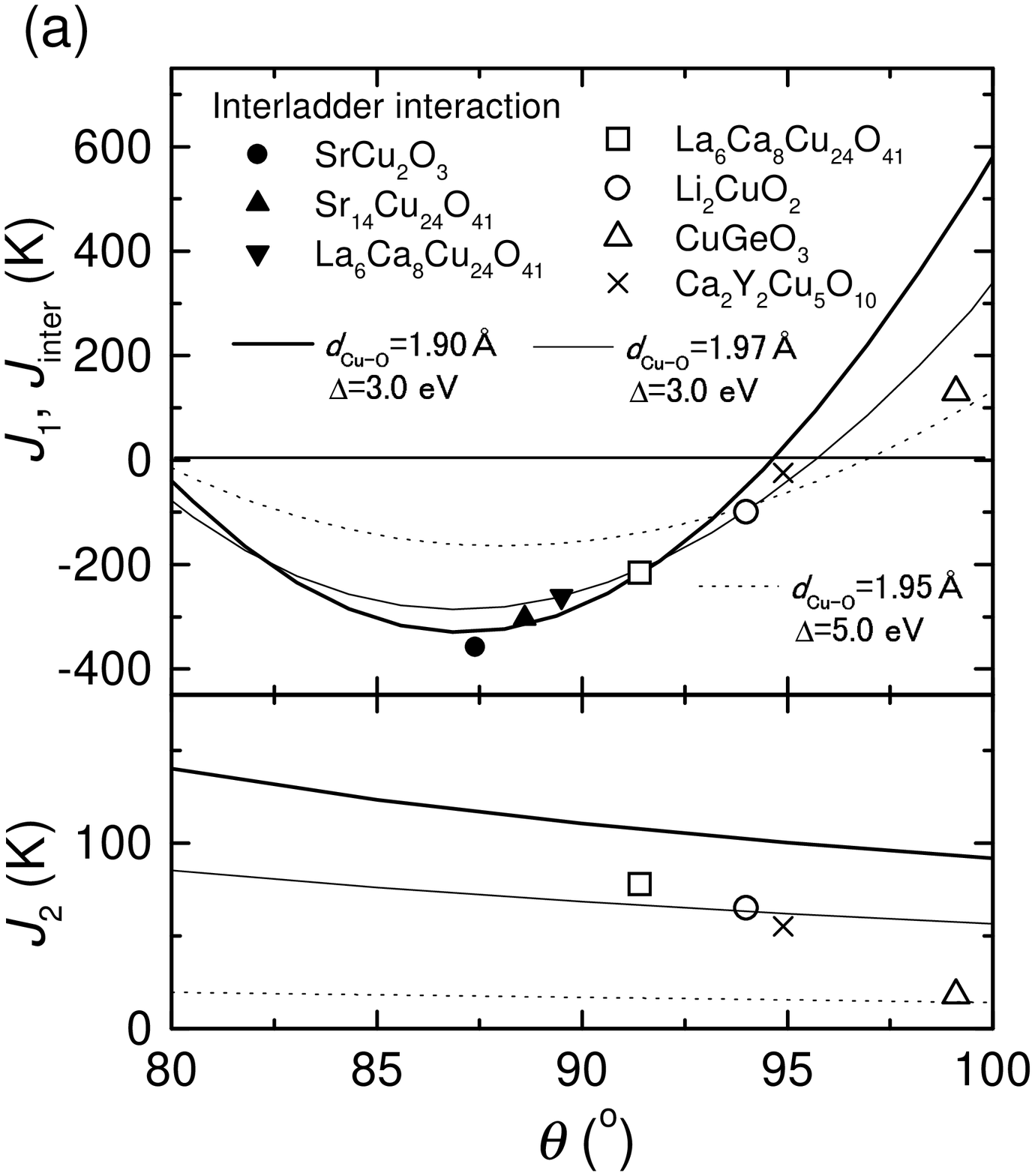,scale=0.4}
\epsfile{file=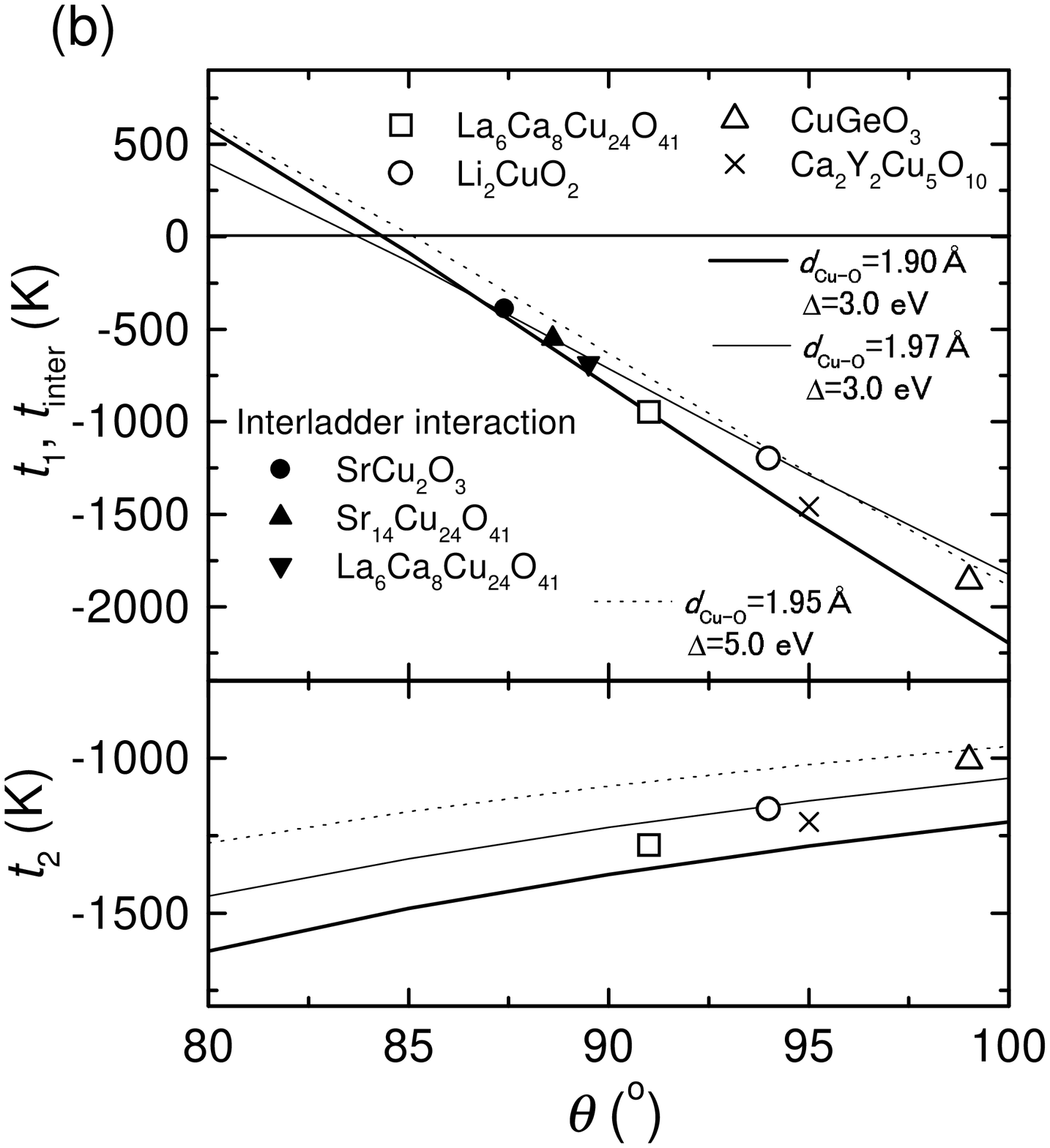,scale=0.4}
\caption{(a) The dependences of $J_1$, $J_2$ and $J_{\rm inter}$ on $\theta$ for several choices of parameter sets of $d_{\rm Cu-O}$ and $\Delta$. (b) The dependences of $t_1$, $t_2$ and $t_{\rm inter}$ on $\theta$ for several choices of parameter sets of $d_{\rm Cu-O}$ and $\Delta$. The squares, the circles and the triangles denote $J_1$ and $J_2$  for La$_6$Ca$_8$Cu$_{24}$O$_{41}$, Li$_2$CuO$_2$ and CuGeO$_3$, respectively.}
\label{fig2}
\end{center}
\end{figure}

In summary, we have discussed the hole distribution of Sr$_{\rm 14-x}$Ca$_{\rm x}$Cu$_{24}$O$_{41}$ from the viewpoint on  hole doping dependences of $\sigma(\omega)$ for the edge-sharing chain. When holes are introduced, a new structure, the intensity of which increases with hole doping, appears at around 3~eV in $\sigma(\omega)$. From the comparison with the experiments, We conclude that Ca doping moves the holes from the chains to the ladders. The strengths of the magnetic interactions and the hopping energies of the local singlet have been obtained by the the small cluster model including the realistic interactions. The edge-sharing chain are expected to show the various the excitation spectra, depending on the bond angle of Cu-O-Cu. The further study of the excited spectra on the effective $t_1$-$t_2$-$J_1$-$J_2$ model are now in progress.

This work was supported by a Grant-in-Aid for Scientific Research on Priority Areas from the Ministry of Education, Science, Sports and Culture of Japan. The parts of the numerical calculation were performed in the Supercomputer Center, ISSP, University of Tokyo,  and the supercomputing facilities in IMR, Tohoku University.  Y. M. acknowledges the financial support of JSPS Research Fellowships for Young Scientists.


\begin{thebibliography}{99}
\bibitem{ef} The $t_1$-$t_2$ part of the $t_1$-$t_2$-$J_1$-$J_2$ Hamiltonian is defined as $H_{t_1,t_2}=-t_1 \sum_{i}c_{i}^{\dagger}c_{i+1}-t_2 \sum_{i}c_{i}^{\dagger}c_{i+2}$.
\end{thebibliography}
\end{document}